\documentclass[aps,prl,nofootinbib,twocolumn,superscriptaddress]{revtex4}
\pdfoutput=1

\usepackage{graphicx}
\usepackage{amsmath}
\usepackage{hyperref,color}

\newcommand{\be}{\begin{equation}}
\newcommand{\ee}{\end{equation}}
\newcommand{\br}{\begin{eqnarray}}
\newcommand{\bea}{\begin{eqnarray}}
\newcommand{\eea}{\end{eqnarray}}
\newcommand{\er}{\end{eqnarray}}
\newcommand{\ba}{\begin{array}}
\newcommand{\ea}{\end{array}}
\newcommand{\bi}{\begin{itemize}}
\newcommand{\ei}{\end{itemize}}
\newcommand{\bn}{\begin{enumerate}}
\newcommand{\en}{\end{enumerate}}
\newcommand{\bc}{\begin{center}}
\newcommand{\ec}{\end{center}}

\newcommand{\eq}[1]{eq.~(\ref{#1})}

\newcommand{\beq}{\begin{equation}}
\newcommand{\eeq}{\end{equation}}

\newcommand{\gsim}{\lower.7ex\hbox{$\;\stackrel{\textstyle>}{\sim}\;$}}
\newcommand{\lsim}{\lower.7ex\hbox{$\;\stackrel{\textstyle<}{\sim}\;$}}

\def\mysection#1{\bigskip\paragraph{\bf #1.} }

\begin{document}

\title{The cosmic-ray positron excess from a local Dark Matter over-density 
 }

\author{Andi Hektor}
\affiliation{National Institute of Chemical Physics and Biophysics, Ravala 10, 10143 Tallinn, Estonia}
\affiliation{Helsinki Institute of Physics, P.O. Box 64, FI-00014, Helsinki, Finland}
\author{Martti Raidal}
\affiliation{National Institute of Chemical Physics and Biophysics, Ravala 10, 10143 Tallinn, Estonia}
\affiliation{Institute of Physics, University of Tartu, Estonia}
\author{Alessandro Strumia}
\affiliation{National Institute of Chemical Physics and Biophysics, Ravala 10, 10143 Tallinn, Estonia}
\affiliation{Dipartimento di Fisica dell'Universit{\`a} di Pisa and INFN, Italia}
\author{Elmo Tempel}
\affiliation{National Institute of Chemical Physics and Biophysics, Ravala 10, 10143 Tallinn, Estonia}
\affiliation{Tartu Observatory, Observatooriumi 1, 61602 Toravere, Estonia}

\date{\today}

\begin{abstract}
We show that the cosmic-ray positron excess measured by PAMELA and AMS 
could be induced by Dark Matter annihilations in a local over-density.
In such a context leptophilic DM is not needed and good fits to positron data, 
in agreement with antiproton and gamma-ray measurements,
are obtained for 
DM annihilations to $WW$, $hh$, $ZZ$, $ t \bar t$,  $b \bar b$, $q\bar q$  channels.
The classic Dark Matter candidates, such as the pure supersymmetric Wino
with standard thermal annihilation cross-section, can fit the positron excess,
without invoking any additional  assumption on Dark Matter properties.

\end{abstract}


\maketitle

\mysection{Introduction}
The new AMS02 measurement~\cite{ams,ams2} of the cosmic ray positron energy spectrum up to 350~GeV 
confirms with better precision the earlier claim by PAMELA~\cite{FermiLAT:2011ab} and FERMI~\cite{Adriani:2008zr}  
of a rising  positron/electron fraction.
Such a spectral feature demands either non-conventional models of the astrophysical background~\cite{Blasi}
or new sources, such as pulsars~\cite{Hooper:2008kg,Malyshev:2009tw,Barger:2009yt,Grasso:2009ma,Linden:2013mqa,Cholis:2013psa} or 
annihilations of weakly interacting  Dark Matter (DM).


DM can explain the positron excess 
compatibly with the absence of a similar excess in the antiproton flux
provided that the DM of the main Milky Way halo  annihilates predominantly into the Standard Model (SM)
leptons with a cross-section $2-3$ orders of magnitude larger than the  annihilation cross-section 
predicted by the hypothesis that DM is a thermal relic~\cite{Cirelli:2008pk,ArkaniHamed:2008qn,Cholis:2008wq}. 
Such a large cross-section today may result from a Sommerfeld enhancement~\cite{Cirelli:2008pk}, maybe  
mediated by new hypothetical GeV scale vectors~\cite{ArkaniHamed:2008qn}.
However, this scenario is severely constrained by
the absence of associated gamma-rays from the galactic center, from dwarf galaxies and in the diffuse background~\cite{Meade:2009iu,Cirelli:2009dv,Papucci:2009gd,Huetsi:2009ex,Ackermann:2011wa,Ackermann:2012rg}. 
Additional constraints arise from observations of the cosmic microwave background (CMB)~\cite{Galli:2009zc,Slatyer:2009yq,Huetsi:2009ex,Cirelli:2009bb}.
Such constraints challenge various aspects of
current DM theories  --- DM origin as a thermal relic, early cosmology, simulations  
of the Milky Way DM halo density profile, as well as particle physics models of the DM.


Furthermore, even DM annihilations into leptons are challenged, because the final state $e^\pm$ loose almost all of their energy
through inverse Compton scattering on galactic star-light and CMB, producing a secondary flux of energetic photons.
Such Inverse Compton photons can be compatible with
 gamma-ray observations provided that DM in the Milky Way has a cored (such as an isothermal) density profile~\cite{us,Kopp:2013eka,DeSimone:2013fia,Yuan:2013eja,Jin:2013nta,Yuan:2013eba,Ibe:2013nka,Kajiyama:2013dba}.

The non-observation of such Inverse Compton photons favors the possibility that the positron excess is local,
rather than present in all the Milky Way.

\medskip

In this paper we propose a  solution to the positron anomaly that does not require additional
{\it ad hoc} assumptions on DM properties. The idea is that the positron anomaly is a local effect
arising from DM annihilations in a local DM over-density.  DM density fluctuations, that are not gravitationally bound, 
 are predicted to occur and disappear  continuously everywhere in our Galaxy by the  cold DM paradigm.  The measured positron excess could 
then originate from such a local over-density even with the standard thermal annihilation cross-section.\footnote{Various past articles
considered the possibility of interpreting the positron excess in terms of enhanced DM matter
annihilations from a variety of different kinds of nearby DM sub-structures, such as
clumped sub-haloes~\cite{Hooper:2008kv}, black holes~\cite{Saito:2010ts}, dark stars~\cite{Ilie:2010vg}, a dark disk~\cite{Cholis:2010px}.} 

This implies observable energy spectra of $e^\pm,\bar p,\gamma$ different from the standard case where DM annihilates
in all the Galaxy.
Our most important result is that  DM annihilations into the usual theoretically favoured channels,
$$WW,\ ZZ,\ hh,\ q\bar q,\ b\bar b,\ t\bar t,\ \ldots,$$
can now reproduce the energy spectrum of the positron excess, 
while purely leptonic channels become disfavoured. 
This is because positron energy losses can now be neglected, such that
a more shallow energy spectrum at production is needed to fit the positron excess.
This result implies that the conventional WIMP DM models are preferred by data without invoking additional assumptions. 
We will show that constraints from $\bar p$ and $\gamma$ are satisfied.

Additional information that may discriminate between DM models is provided by DM direct detection experiments.
If the local DM over-density exists today around us, the DM coupling to nuclei must be suppressed. This favours, for example, pure
 Wino DM or Minimal Dark Matter scenarios, where DM couples to matter only via a $W$-boson loop.
If, instead, the DM over-density has already disappeared, and today we observe a remnant position excess trapped by the Galactic magnetic fields,
 typical WIMP DM model are viable candidates.

\mysection{The local DM over-density}
$N$-body simulations of cold DM structure formation predict a wide spectrum of density and velocity fluctuations in any DM halo such as our Galaxy~\cite{Nfl}.
Only a very small fraction of the density fluctuations develop high enough over-density, a few hundred times over the local average,  
to become gravitationally bound sub-halos. 
A fluctuation with density $\rho$ and radius $R$
is  gravitationally bound provided that the escape velocity from it is smaller than the typical local DM velocity dispersion,
\beq 
v_{\rm esc} = \sqrt{\frac{8\pi }{3}G_N R^2 \rho } \ll 10^{-3},
\eeq
i.e.\ for $\rho/\rho_0 \gg 200 \times  (\hbox{kpc}/R)^2$,
where $\rho_0 = 0.4$~$\rm{GeV/cm^3}$ is the average DM density around the Sun.
Those dense fluctuations collapse gravitationally and develop cuspy NFW or Einasto like profile similarly to the main halo.
However, the vast majority of fluctuations just occur and disappear continuously without affecting large scale structure formation. Those over-density regions
have shallow profiles, such as Gaussians, since they are not gravitationally bound.

In this work we assume that there exists, or there existed not long time ago, a local DM over-density with a radius of few hundred pc. 
Such an assumption is in agreement with the determination of local DM density at the distance of Sun.
The latter is measured by the movement of stars in a cylinder of radius 1~kpc extending $\pm 4$~kpc in both directions around the Galactic disc.
A local over-density with radius $R=100$~pc forms just 1/6000 of this volume, not affecting the average result. In fact, several over- and under-density
fluctuations are expected to occur in such a big volume.
Furthermore, a moderate local over-density is compatible with
solar system gravitational measurements that imply a local DM density smaller than  $\rho/\rho_0 < 15000$~\cite{Pitjev}.

\mysection{Explaining the positron excess}
We now try to interpret the measured positron excess as due to
DM annihilations in a local DM over-density.
As a result, we obtain the size and density of such a fluctuation from the AMS and PAMELA data. This will allow us to later {predict} the 
associated gamma-ray and antiproton fluxes.

In our exploration we follow the model independent approach introduced  in~\cite{Cirelli:2008pk,Cirelli:2010xx}.
We allow DM to annihilate into all possible two-body SM final states with the standard thermal relic cross-section 
$\langle \sigma v \rangle \approx 3\cdot 10^{-26}$~cm$^2.$
The energy spectra of the various stable SM particles ($e^+,\bar p,\gamma,\nu,\ldots$) are computed
with PYTHIA8. 

To compute the diffusion effects we assume the MIN/MED/MAX diffusion models of~\cite{minmedmax},
as described in \cite{Cirelli:2010xx}.  
The number densities $f(\vec x, t,E)$ of $e^+$ and their fluxes $\Phi_{e^+} = c f/4\pi$ are well approximated by neglecting the
energy loss term in the time-independent  diffusion equation, that becomes simply
\beq -K(E) \nabla^2 f = Q \frac{dN_{e^+}}{dE}= \frac12 \frac{\rho^2}{M^2}  \langle \sigma v \rangle   \frac{dN_{e^+}}{dE},
\eeq
where $K(E)$ is approximatively given by the Larmor radius in the local turbulent Milky Way magnetic field.
Analyses of cosmic ray data suggest
$K(E) = K_0 (E/{\rm GeV})^\delta$ with 
$\delta=0.85- 0.46$ and $K_0 = (0.016 - 0.0765)\,{\rm kpc}^2/{\rm Myr}$~\cite{minmedmax}.

\begin{figure}[t]
\begin{center}
\includegraphics[width=0.45\textwidth]{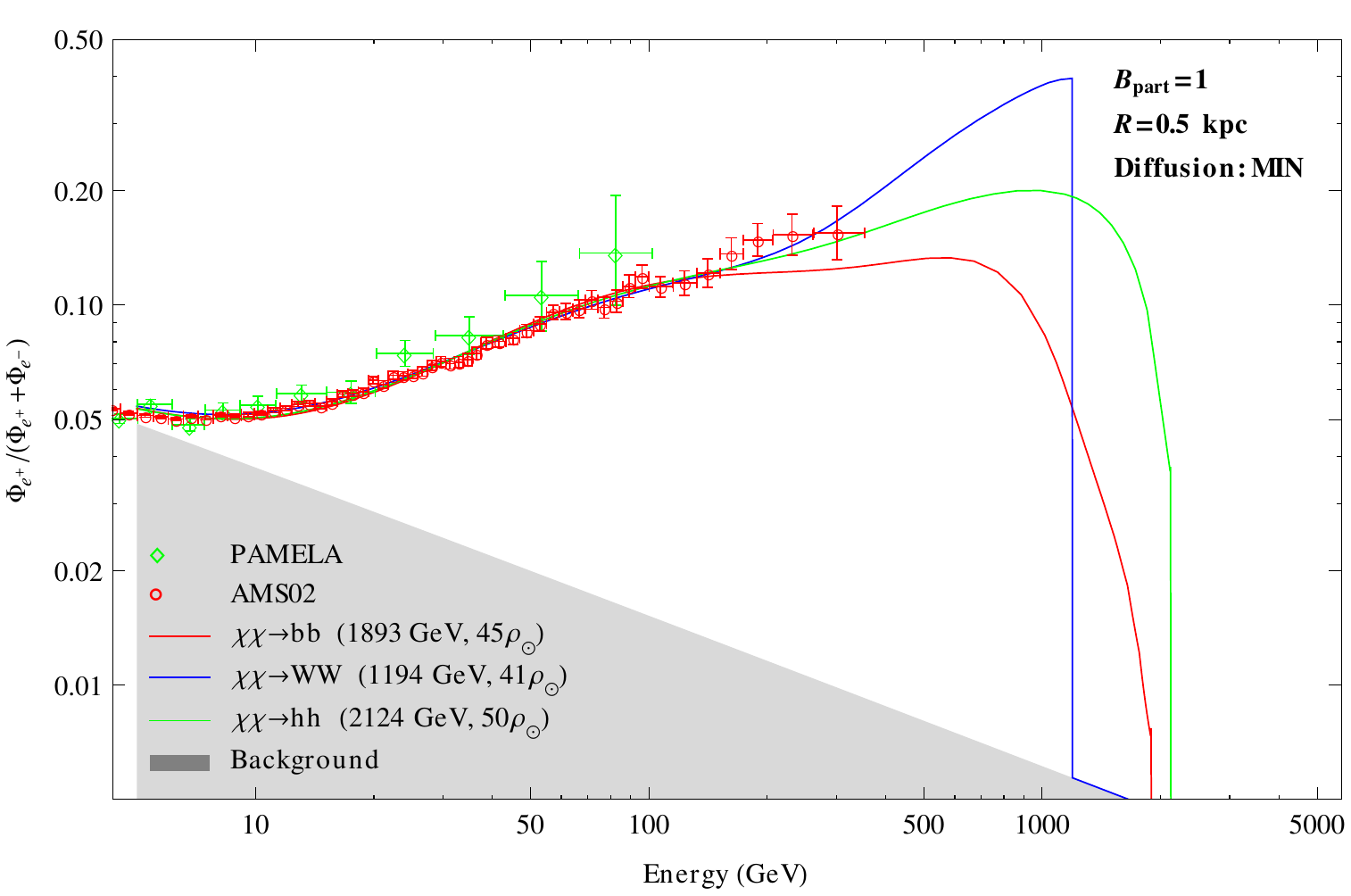}
\caption{\em Best fits of the positron fraction from DM annihilations to $WW$, $hh$,  $\bar b b$  for parameters indicated in the figure.
Over-densities are indicated as $\rho_{\rm loc}$ and given relative to the average density $\rho_0$.} 
\label{fig1}
\end{center}
\end{figure}

Assuming, for simplicity, that we live at the center of a spherical excess with constant local density $\rho$ and radius $R$,
and neglecting DM annihilations outside it, the solution is
\beq \label{eq:Phi}
 \Phi_{e^+} = 
\frac{3\Gamma}{32\pi^2 K(E) R}\frac{dN_{e^+}}{dE},\eeq
where $\Gamma$ is the total DM annihilation rate in the local over-density:
\beq \Gamma =
\int dV\, Q= \frac{4\pi R^3}{3}   \langle \sigma v \rangle \frac{1}{2}\frac{\rho^2}{M^2} .
\eeq
The shape and location of the local excess only affect the overall numerical factor in eq.~(\ref{eq:Phi}),
leaving unaffected the main feature: {\em the positron energy spectrum at detection is given by the positron energy spectrum at production
over the diffusion factor $K(E)$}.

\bigskip

The boost factor $B$ that enhances the positron DM flux with respect to the standard scenario 
can be expressed as
\bea
B=B_{\rm part} \times B_{\rm local},
\label{boost}
\eea
where $B_{\rm local}\propto \rho^2$  is the boost induced by the local DM over-density that we are considering, and
$B_{\rm part}$ is a possible extra boost due to particle physics, not needed in our context,
but that could be anyhow present.
For example, DM with SM weak interactions and heavier 
than $M_{\rm DM}> M_W/\alpha_2\approx \hbox{few TeV}$
has an annihilation cross section enhanced at low velocity by
the electroweak SM Sommerfeld effect, thereby producing $B_{\rm part}>1$.


\begin{figure}[t]
\begin{center}
\includegraphics[width=0.5\textwidth]{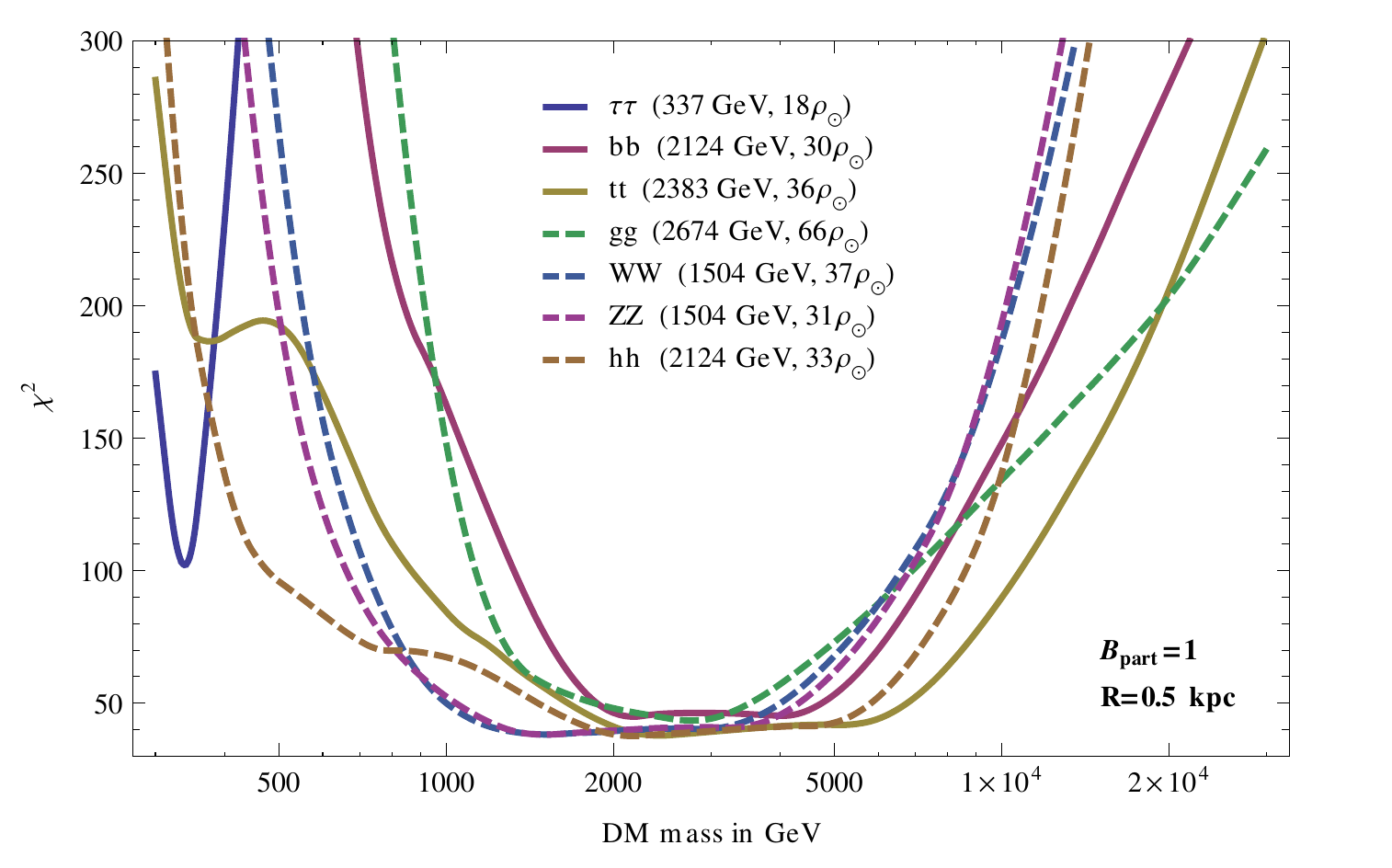}
\caption{\em  Fit to the positron fraction: $\chi^2$ as function of the DM mass
for different DM annihilation channels. 
\label{fig2}}
\end{center}
\end{figure}

\medskip

In Fig.~\ref{fig1} we plot the best fit spectra of the positron fraction from the DM annihilations to $WW$, $hh$ and  $\bar b b$ channels
 as functions of positron energy. We assumed a spherical local over-density with radius $R=500~{\rm pc}$.
 The $\chi^2$ of the fits for the various annihilation channels as function of the DM mass
 are presented in Fig.~\ref{fig2}.  The required over-densities are also presented in the figures. 

The main result from Figs.~\ref{fig1}, \ref{fig2} is that only DM annihilations to channels like $WW$, $ZZ,$ $gg$, $hh$,  $\bar b b$, $\bar t t$ give good fits to data
while leptonic channels give very poor ($e^+e^-,\mu^+\mu^-$) or poor ($\tau^+\tau^-$) fits.

The reason for this result is that in our scenario the positron anomaly is  a local phenomenon so that positron energy losses can be neglected.
Therefore,  the measured rise of the positron fraction is reproduced by injecting a  shallow initial positron spectrum $dN_{e^+}/dE$
into the Galactic environment.
This is exactly opposite to the scenario in which the positron excess arises from DM annihilations in the main Galactic halo thanks to a large particle physics boost factor  $B_{\rm part}\gg 1.$
In the latter case the positron energy loss effects are significant and the injected spectrum must be hard to fit data~\cite{us}. This is the reason why only leptonic channels are
able to provide good fits in that case~\cite{us}. Therefore, particle physics models that are able to fit the data are completely different in the two cases.

AMS data prefer DM with masses $1-5$ TeV. As seen in Fig.~\ref{fig1}, 
the high energy behaviour of different annihilation channels are different. 
Measurements of the positron fraction at higher energy will 
provide more informations on the properties of DM.

Notice also that the required DM over-densities for the best fit channels are of order $40-50$: smaller than 
the over-density that would form gravitationally bound sub-haloes.  In presence of particle boost factor $B_{\rm part}$ of order 10, the needed over-density would be reduced down to $\rho \sim 5 \rho_0$.

Based on this scenario, one expects a directional asymmetry of the positron signal,
at the level or smaller than the asymmetry produced by nearby pulsars or  DM sub-haloes~\cite{Linden:2013mqa},
and thereby compatible with existing data.  Given that we do not know the location of the local DM excess relative to us,
such asymmetry cannot be precisely predicted.

Furthermore, the positron excess should also be visible as a small bump in the $e^++e^-$ cosmic ray energy spectrum.
The experimental situation is at the moment unclear: the recent measurement from AMS~\cite{ams2} contradicts earlier measurements from ATIC and FERMI, that contained two different hints of bumps.  Thereby we cannot derive any safe conclusion from present data.

\begin{figure}[t]
\begin{center}
\includegraphics[width=0.45\textwidth]{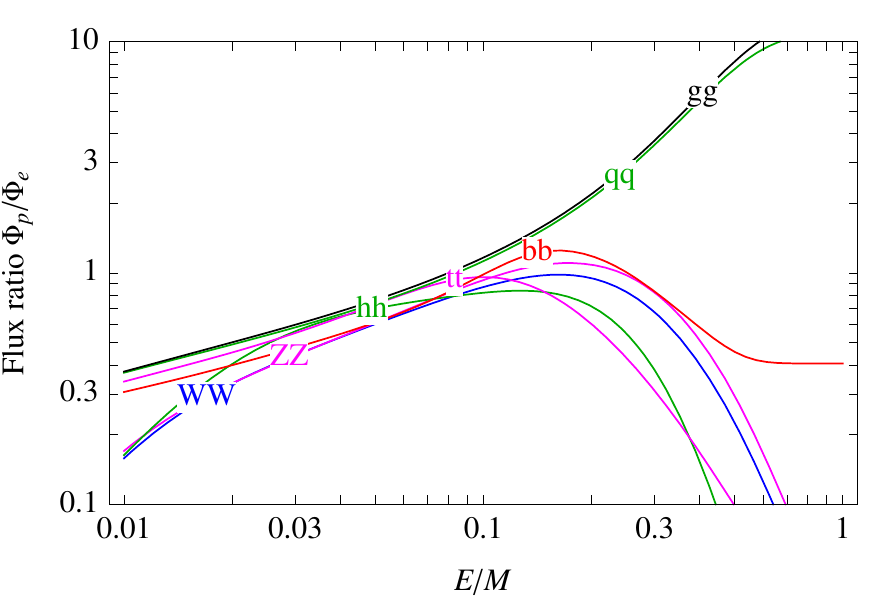}
\caption{\em Predicted $\Phi_{\bar p}/\Phi_{e^+}$ for various SM annihilation channels into
$W^+ W^-, ZZ, hh, t\bar t, b\bar b, q\bar q, gg$ for $M\sim 1\,{\rm TeV}$.} 
\label{dNratio}
\end{center}
\end{figure}

\mysection{Implications for  antiprotons}
Fixing the local DM over-density and the DM parameters as in Figs.~\ref{fig1}, \ref{fig2}, we are able to 
{\it predict} the antiproton fluxes from the local over-density due to
DM annihilations.

In the relevant energy range $\bar p$ and $e^+$ diffuse in the same way, because
they have the same electric charge (up to the sign), because they are both ultra-relativistic, and
because we can neglect positron energy losses and $\bar p$ interactions. 
The $\bar p$ flux is then given by eq.~(\ref{eq:Phi}), just inserting the appropriate prompt energy spectrum.
The prediction is:
\beq\label{PhiR} \frac{\Phi_{\bar p}}{\Phi_{e^+}} = \frac{dN_{\bar p}/dE}{dN_{e^+}/dE}.\eeq
All non-leptonic SM annihilation channels predict that this ratio is $\approx 0.5$ at the relevant value of $E/M \approx 0.1$,
see Fig.~\ref{dNratio}.

\begin{figure}[t]
\begin{center}
\includegraphics[width=0.45\textwidth]{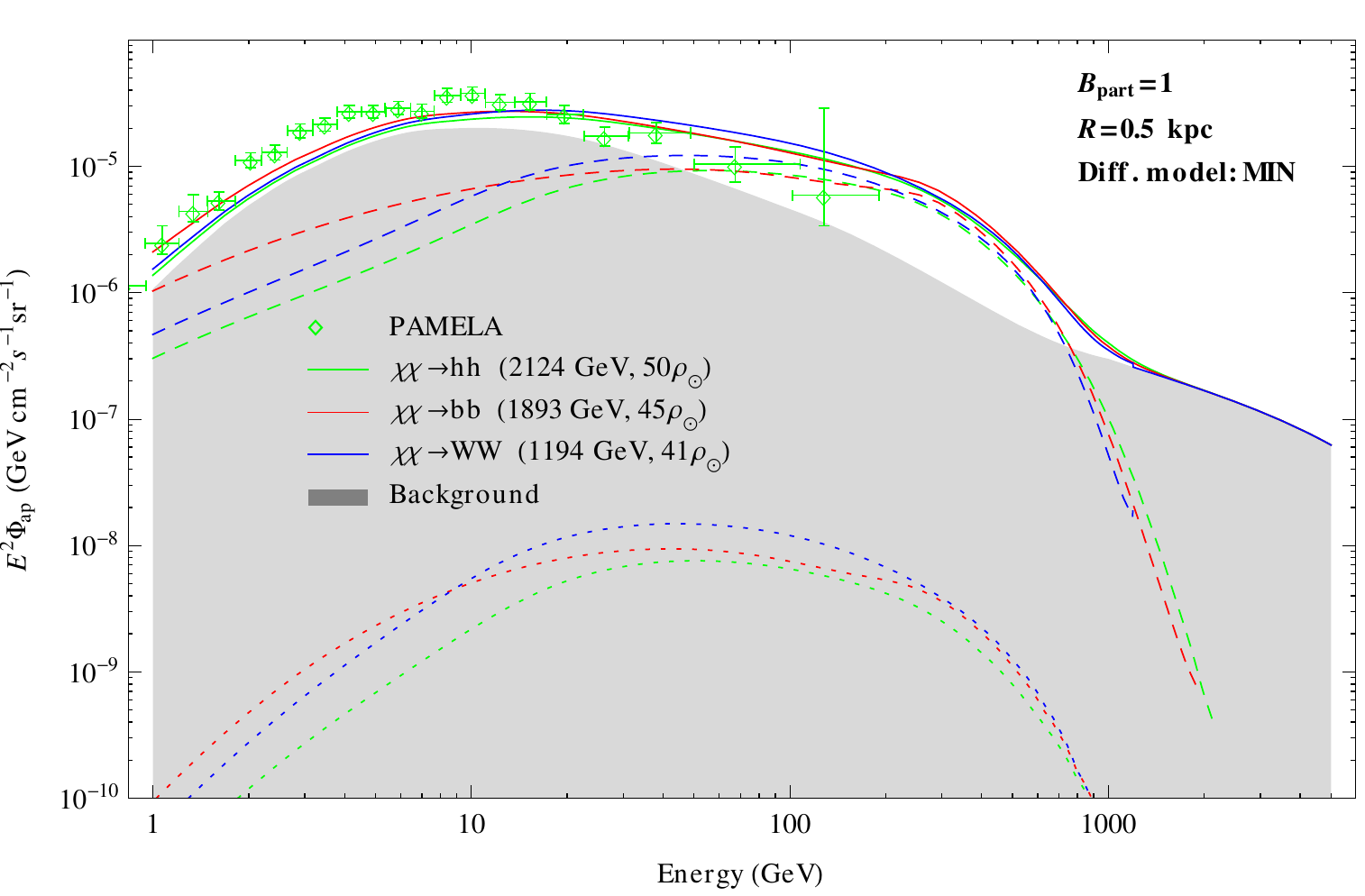}
\caption{\em Predicted $\bar p$  fluxes from the local DM over-density (long dashed) and from the main halo (short dashed) 
for the parameters that in Fig.~\ref{fig1} provide the best fits to the $e^+$ excess.
We also show the estimated astrophysical $\bar p$ background.
} 
\label{fig4}
\end{center}
\end{figure}

This implies a predicted $\bar p$ DM flux at the level of the flux observed by PAMELA, as presented in Fig.~\ref{fig4}.

The grey area in Fig.~\ref{fig4} is the antiproton astrophysical background, as estimated in~\cite{cirelli}.
Given that the astrophysical $\bar p$ background is believed to have a $\sim 30\%$ uncertainty~\cite{grasso},
there is some tension with the PAMELA data at higher energy.  
The issue will be soon  clarified by improved AMS measurements of the $\bar p$ flux.

\bigskip

For comparison,   the standard scenario without a local DM over-density
predicts a  $\Phi_{\bar p}/\Phi_{e^+}$ which is {\em uncertain and
higher} than in \eq{PhiR}, because
energy losses from distant DM annihilations around the center of the Galaxy
reduce $\Phi_{e+}$ but not $\Phi_{\bar p}$,
and because the amount of $\Phi_{\bar p}$ that reaches us depends on the unknown volume of the Galactic diffusion region.
This is why, in the standard scenario, $\bar p$ bounds are stronger and one
needs to select leptophilic DM particle physics models that avoid annihilations into $\bar p$.

\mysection{Implications for  gamma-rays}
The $\gamma$ flux is predicted as
\beq 
\label{gammaflux}
\Phi_\gamma
=
\frac{\langle\sigma v\rangle}{4\pi} 
   \frac{dN_\gamma}{dE} 
\int_{\rm l.o.s.}ds\  \frac{1}{2} \frac{\rho^2}{M^2}   = \frac{3\Gamma}{16\pi^2 R^2}\frac{dN_\gamma}{dE} ,
\eeq
where in the last expression we evaluated the line-of-sight integral by assuming again that we are at the center of
a spherical constant over-density with radius $R$.
The ratio between photons and positrons
is predicted as
\beq \frac{\Phi_\gamma}{\Phi_{e^+}} = \frac{ 2K}{R}\frac{dN_{\gamma}/dE}{dN_{e^+}/dE}\ .\eeq
Up to the geometry-dependent order one factor,
the  astrophysical factor $K/R$ can be intuitively understood as follows.
For all particles, fluxes are inversionally proportional to their speed.
Photons travel at the speed of light, while $e^+$ diffuse in a time $T$ for a distance $R$ with an average velocity
given by $R/T \simeq \sqrt{KT}/T \simeq K/R$.

\begin{figure}[t]
\begin{center}
\includegraphics[width=0.45\textwidth]{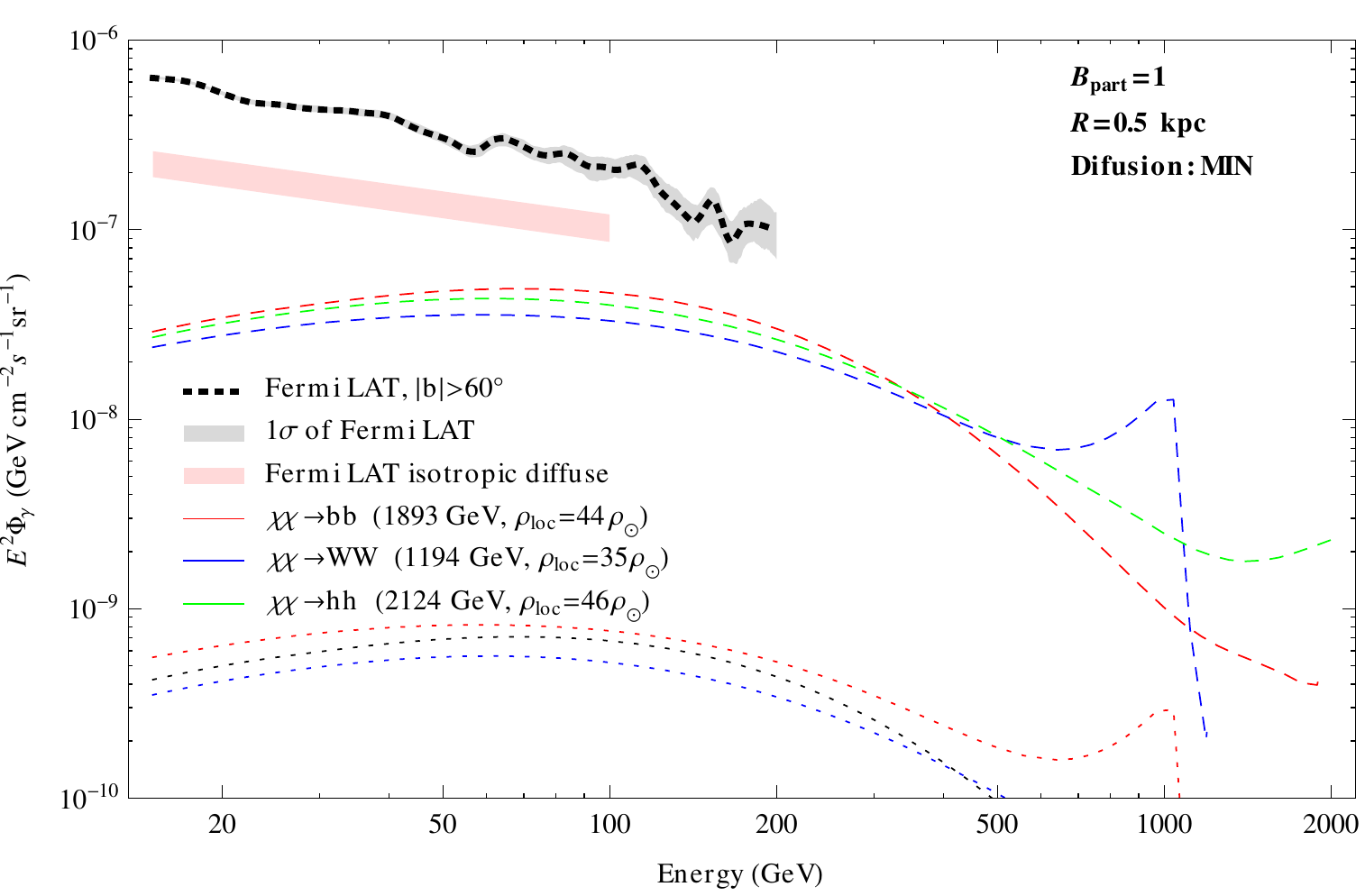}
\caption{\em Predicted gamma-ray fluxes from the local over-density (long-dashed) and the main halo (short dashed) for the same parameters as in Fig~\ref{fig1}. The bands show $\gamma$-measurements: (gray) $\gamma$-ray flux from the polar regions ($|b|>60^{\circ}$) measured by Fermi LAT and (pink) the isotropic component of $\gamma$-ray sky estimated by the Fermi LAT Collaboration.}
\label{fig3}
\end{center}
\end{figure}

The ratio $\Phi_\gamma/\Phi_{e^+}$ depends on the uncertain diffusion parameter $K(E)$ and on the size of the bubble.
Fig.~\ref{fig3} shows the predicted DM $\gamma$ flux for a bubble with $R \approx 0.5\,{\rm kpc}$ and for the MIN
propagation model.
The $\gamma$ flux from local DM annihilations is a factor of few below the two measured $\gamma$ fluxes plotted in Fig.~\ref{fig3}:
\begin{enumerate}
\item The pink band is the diffuse isotropic $\gamma$-ray background, as extracted from the FERMI collaboration~\cite{FERMIiso}.

\item The slightly higher gray band is extracted by us from FERMI data, following a simpler procedure.
We subtracted known point-like sources and reduced the
Galactic gamma-ray background by restricting the observation region to  high Galactic latitudes, $|b|>60^\circ $.


\end{enumerate}
We do not show the expected astrophysical $\gamma$-ray background
because we do not know any reliable estimate of it.

We here neglected the Inverse Compton $\gamma$ ray flux, 
because it is strongly reduced with respect to the standard scenario,
where it is problematic, by our assumptions that the $e^+$ excess is just local.

Finally, we point out that,
while the main features of our results have been explained with simple approximations,
our numerical results have been derived from a full numerical study where we have taken into account
energy losses for $e^\pm$ and other small effects.
In Figs.~\ref{fig4} and \ref{fig3} we also plotted the
contributions to the gamma-ray and antiproton fluxes coming from regions of the Milky Way outside
from the dominant local over-density.   We see that such contribution is so small that the analysis
would remain unchanged in presence of a moderate $B_{\rm part}\sim 10$, or even larger.



\mysection{Implications for DM direct detection}
We found that the positron excess can be reproduced as due to  DM annihilation with the standard thermal-relict cross section,
assuming a local DM over-density with $\rho \sim 40\rho_0/B_{\rm part}$
(see figures~\ref{fig1}, \ref{fig2}).  Here, $B_{\rm part}\ge 1$ is a boost factor of particle physics origin (e.g.\ Sommerfeld enhancement),
that could be larger than one even if this is not needed in our scenario.

The boost of indirect DM detection signals, proportional to $\rho^2$, is accompanied by
a smaller boost of DM direct detection signals, proportional to $\rho$.
In order to explain the negative 
DM direct searches in {\sc Xenon100}~\cite{Aprile:2012nq}, the DM spin-independent cross-section to nuclei must be smaller than about
$10^{-45}\,{\rm cm}^2$ for $M\sim \,{\rm TeV}$.
This happens {\it naturally} in various  theoretically motivated DM models.

For example, if  DM is a pure supersymmetric Wino
(or, equivalently, a Minimal Dark Matter fermion triplet), the DM relic abundance fixes its mass to be 2.5~TeV.
Such a DM candidate gives a good fit to the position excess, as seen in Fig.~\ref{fig1}.
At the same time, such particle couples to nuclei only via a $W$-boson loop, 
giving a small cross section
$\sigma_{\rm SI}\sim 0.6~10^{-46}\,{\rm cm}^2$~\cite{Hisano:2010fy}, compatible with 
the negative results of {\sc Xenon100}.

Alternatively, in many models DM couples to SM particles only via the Higgs doublet.
Such models generically predict  DM annihilations into $hh$ and may have small enough DM/nucleon cross section $\sigma_{\rm SI}$. 
In particular, scenarios in which the electroweak breaking scale is  induced via the Higgs boson mixing with a singlet scalar from 
the dark sector~\cite{Heikinheimo:2013fta,Heikinheimo:2013xua,natHS}
predict generically that  $\sigma_{\rm SI}$ is suppressed by the small mixing angle.

\medskip


If, instead, the local DM over-density fluctuation has already disappeared today, and PAMELA, AMS measure 
the remnant of the positron excess trapped by Galactic magnetic fields, no additional constraints 
on our scenario occurs from DM direct detection experiments.

\mysection{Conclusions}
We have shown that the positron excess measured by PAMELA and confirmed by AMS
could be due to DM annihilations enhanced by a local DM over-density.
In such a context, it is not necessary to assume leptophilic DM annihilations ---
on the contrary DM annihilations into the theoretically favoured channels
 $WW$, $ZZ,$ $gg$, $hh$,  $\bar b b$, $\bar t t$  can explain the data.
This scenario predicts $\bar p$ and $\gamma$ fluxes from DM annihilations at the level of present measurements.
In particular, AMS can test this scenario performing and improved measurement of the $\bar p$ flux.
In such a context, the positron excess
prefers `classical' WIMP DM candidates with suppressed coupling to nuclei, such as the pure Wino, without
additional assumptions on DM properties nor invoking any exotic particle physics to boost the annihilation cross-section.

\medskip

Finally, if the positron excess is not due to DM annihilations,
our results imply a bound on the local
DM density that is stronger than the direct bound~\cite{Pitjev}
for $M\sim$ 1 TeV and for a radius $R> 0.1\,{\rm pc}$ of the local over-density,
and under the assumption of a thermal DM annihilation cross section.

\mysection{Acknowledgement}
AH thanks Peter H. Johansson for very helpful discussions. This work was supported by the ESF grants 8499, 8943, MTT8, MTT59, MTT60, MJD52, MJD140, JD164, MJD298, MJD272 by the recurrent financing SF0690030s09 project and by the European Union through the European Regional Development Fund.

\end{document}